\documentclass[conference]{IEEEtran}

\usepackage[T1]{fontenc}
\usepackage{hyperref}
\usepackage{url}
\usepackage{graphicx}
\usepackage{adjustbox}
\usepackage{amssymb, amsmath}
\usepackage{amsfonts}
\usepackage{makecell} 
\usepackage{cite}
\usepackage{bm}
\usepackage{cases}
\usepackage[table, dvipsnames]{xcolor}
\usepackage{enumitem}
\usepackage{algpseudocode}
\usepackage{algcompatible}
\usepackage{algorithm}
\usepackage{caption}
\usepackage{booktabs}
\usepackage{multirow}
\usepackage{physics}
\usepackage{multicol}
\usepackage{float}
\usepackage{subcaption}
\usepackage{booktabs,dcolumn}
\floatstyle{ruled}
\usepackage{color,soul}
\newfloat{model}{thp}{lop}
\floatname{model}{Model}

\usepackage{multirow,mdframed}
\mdfsetup{skipabove=0pt,skipbelow=0pt}
\mdfdefinestyle{model}{%
  innertopmargin=0pt,
  innerbottommargin=0pt,
  innerleftmargin=0pt,
  innerrightmargin=0pt,
  skipbelow=0pt,%
  skipabove=0pt,
  splitbottomskip=0pt,
  splittopskip=0pt,
  leftmargin =0pt,% 
    rightmargin=0pt,
  splittopskip=0pt,
    usetwoside=false,
}
\restylefloat{algorithm}
\definecolor{darkgreen}{RGB}{0,135,0}
\definecolor{coral}{RGB}{255,115,80}

\begin{document}

\title{ Bus Type Switching to Reduce \\Bound Violations in AC Power Flow
\thanks{This research is supported by National Science Foundation awards $2041835$ and $2242930$.}}

\author{\IEEEauthorblockN{Anna Van Boven}
\IEEEauthorblockA{\textit{Dept. of Civil, Environmental, and Architectural Engr.} \\
\textit{University of Colorado Boulder}\\
Boulder, CO, USA \\
anna.vanboven@colorado.edu}
\and
\IEEEauthorblockN{Kyri Baker}
\IEEEauthorblockA{\textit{Dept. of Civil, Environmental, and Architectural Engr.} \\
\textit{University of Colorado Boulder}\\
Boulder, CO, USA \\
kyri@colorado.edu}
}
\IEEEoverridecommandlockouts
\maketitle
% \pagenumbering{gobble}

\begin{abstract}
Wholesale power markets often use linear approximations of power system constraints. 
Because it does not consider inequality constraints, using AC power flow for feasibility post-processing can violate bounds on reactive power, voltage magnitudes, or thermal limits. 
There remains a need for a streamlined analytical approach that can guarantee AC feasibility while adhering to variable bounds. This paper suggests an augmented implementation of AC power flow that uses an additional two bus types (PQV and P) to help resolve voltage bound violations present in the traditional approach.
The proposed method sacrifices the voltage setpoint at a generator in exchange for fixing the voltage at a load bus, thereby moving a degree of freedom around the network. Results on the IEEE 14-bus, 57-bus, and 300-bus test cases demonstrate how switching bus types can reduce overall network violations and help find feasible power system setpoints.
\end{abstract}

\begin{IEEEkeywords}
Feasibility, AC Power Flow, Extended Bus Types
\end{IEEEkeywords}

% \maketitle

\section{Introduction}\label{sec: intro}
The power grid is a large system with many non-convex constraints, whose operations have a significant financial impact on electricity producers and individual households. Operating the power grid is therefore a careful balance between economic decision-making and physical feasibility. The power market is often cleared in small time steps (5-15 minute intervals). Since it is computationally challenging to solve large-scale, non-convex optimization problems in this short of a time frame, market operations rely on convex approximations of power grid physics, like DC-Optimal Power Flow (DC-OPF). While useful, these approximations guarantee that the market clearing setpoints are never truly AC feasible \cite{kyri_feas, relax_ac_feas}.

\begin{figure*}[t!] 
\centering \vspace{1mm}
\includegraphics[scale = 0.8]{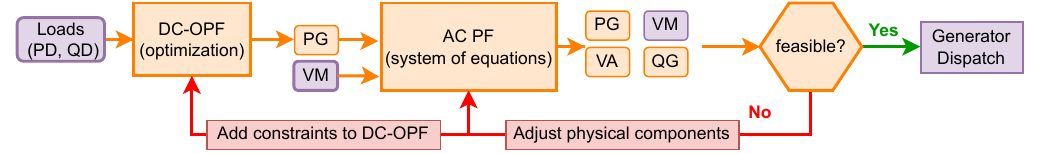} 
\caption{Assumed information flow for power grid operational procedures.} \label{fig:market}\vspace{-4mm} 
\end{figure*}

Power grid operational procedures differ between markets and between countries. One popular procedure (outlined in Figure \ref{fig:market}) involves solving the DC-OPF problem in real-time to obtain real power generation setpoints, then using these setpoints to solve a system of non-linear  AC power flow (AC-PF) equations to determine reactive power generation, voltage angles, and voltage magnitudes \cite{ferc_software}. As depicted in red in Figure \ref{fig:market}, this system of equations can fail to converge, or can converge to an infeasible setpoint. U.S. grid operators currently employ a number of techniques to maintain AC-feasibility in real-time operations, including transferring power between Balancing Areas (BAs), switching transmission lines, adjusting settings on physical components such as transformers or shunt devices, or re-running DC-OPF with tighter constraints to resolve bound violations \cite{ots, nrel_acdc}. There are disadvantages to each of these techniques: real-time changes to BA power transfer setpoints are discouraged, constantly adjusting the settings of physical grid components can significantly reduce their lifetime, and re-running DC-OPF is computationally expensive \cite{device_life}. A significant drawback to all of these techniques is that they require control decisions made by grid operators. 

Research in this field offers faster solutions to maintain AC-feasibility without requiring manual adjustments from grid operators. Here, we present the research from the least to the most representative of current  grid operations. In \cite{acopf, canos}, researchers propose methods to learn the solution to AC-Optimal Power Flow (AC-OPF) given only the system load. These methods do not rely on approximated grid physics, and therefore do not require solving DC-OPF or AC-PF. Researchers in \cite{dcacopf, feas_rest} learn the solution to AC-OPF given both the system load and the generator active power setpoints. These methods are closer to current operations as they solve the DC-OPF, but bypass the need for the AC-PF solutions. The method introduced in \cite{state_est_feas} solves a simple optimization problem to map an approximated setpoint to the nearest AC-feasible setpoint, regardless of the approximation used. In this paper, we propose a computational technique that improves the convergence and feasibility of Newton-Raphson AC-PF (NR-ACPF) solutions. We incorporate two additional bus types (P and PQV) into the traditional NR-ACPF procedure to resolve any reactive power and voltage magnitude violations present in the NR-ACPF solution. Our approach is analytical, so does not introduce the same uncertainty as ML-based methods, and easily integrates into a traditional NR-ACPF solver. Additionally, our approach is guaranteed to reduce the number of manual adjustments required to obtain an AC-feasible solution without deviating from the market clearing generation setpoints.

The remainder of this paper is structured as follows: Section \ref{sec: background} describes the traditional power market operations and the AC-PF equations. Our proposed method is introduced in Section \ref{sec: methodology}, and the results are presented in Section \ref{sec: res}. The paper concludes with discussion of our method and concluding remarks in Section \ref{sec: res}.

\section{Background}\label{sec: background}
This section discusses the current power grid operations and  the disadvantages to this approach.

\subsection{Baseline AC Power Flow with Newton-Raphson}\label{sec: nracpf}
A power system with $n$ buses has $4n$ control variables: each bus has a real power ($P$), reactive power ($Q$), voltage magnitude ($V$), and voltage angle ($\theta$).  One generator is identified as the slack bus $s$, which is a generator bus with a large capacity to make up for losses and other discrepancies introduced by DC-OPF. Assume that the system has $g$ buses with generation, $l$ buses with load, and that $g + l + s = n$ (buses with both generation and load are $g$). At the beginning of the real-time market clearing, real and reactive load ($P$ and $Q$) are provided for all load buses, each generator has an assumed voltage magnitude ($V$), and the voltage angle ($\theta$) at the slack bus is set to 0 (PJM and ERCOT establishes default voltage schedules and bandwidths for each generator \cite{pjm, ercot}).  This results in a system with $2l + g + 2s$ knowns and $2l + 3g + 2s$ unknowns. DC-OPF then solves a convex optimization problem to determine the real power at each generator bus, assuming all voltage magnitudes are 1, lines are lossless, and angle differences between lines are small. With these setpoints, the system now has only two unknown values for each bus: load (PQ) buses have values for $P$ and $Q$, but not $V$ and $\theta$. Generation (PV) buses have values for $P$ and $V$, but not $Q$ or $\theta$. Finally, the slack (V$\theta$) bus has a known $V$ and $\theta$, but unknown $P$ and $Q$. 

AC-PF seeks to find a set of control variables such that power balance is satisfied at each bus, meaning the total real or reactive power produced by a bus is equal to the amount of real or reactive power flowing out of the bus. These equations can be defined for each bus $k$ with neighbors $\mathcal{N}_k$ as: 
% \scalebox{0.85}{%
% $\displaystyle
\begin{align*}
    P_k = V_k^2G_{kk} + V_k\sum_{j \in \mathcal{N}_k}& V_j(G_{kj}\cos(\theta_k - \theta_j) \\
    &+ B_{kj}\sin(\theta_k - \theta_j)) \\ 
    Q_k = V_k^2B_{kk} + V_k\sum_{j \in \mathcal{N}_k}& V_j(-B_{kj}\cos(\theta_k - \theta_j) \\
    &+ G_{kj}\sin(\theta_k - \theta_j))  
\end{align*}
% $
% }
Where $G \in \mathcal{R}^{nxn}$ and $B \in \mathcal{R}^{nxn}$ are the respective real and reactive components of the admittance matrix,
and $G_{ij}$ and $B_{ij}$ are the real and reactive elements of the admittance matrix at row $i$ and column $j$.
% $G_{ij}$ and $B_{ij}$ are the real and reactive admittance from bus $i$ to bus $j$, and $G_{ii}$ and $B_{ii}$ are the real and reactive self-admittance. 
Since we now have $2n$ equations and $2n$ unknowns, the unknowns can be solved for using an iterative method for nonlinear equations.

\subsection{AC Power Flow with Q Limits}\label{sec: qlim}
If Newton-Raphson converges, the solution is guaranteed to meet the real and reactive loads with the generation setpoints obtained in DC-OPF. However, these equations alone cannot guarantee complete AC-feasibility: the solution can (and often does) converge to a point that violates the upper and lower bounds of reactive power ($Q$) of generators, voltage magnitudes ($V$) of loads, or the thermal limits of power lines. MATPOWER (a MATLAB toolbox for power systems analysis) offers a strategy to resolve reactive power violation at a generator by fixing a violated $Q$ to its upper/lower bound and switching the generator bus type from PV to PQ \cite{matpower}. In doing so, the generator's $V$ setpoint is sacrificed in order to guarantee $Q$ feasibility. Each bus still has two unknown values, so the AC-PF equations will still converge. 

\begin{figure}[h!] 
\centering \vspace{1mm}
\includegraphics[scale = 0.7]{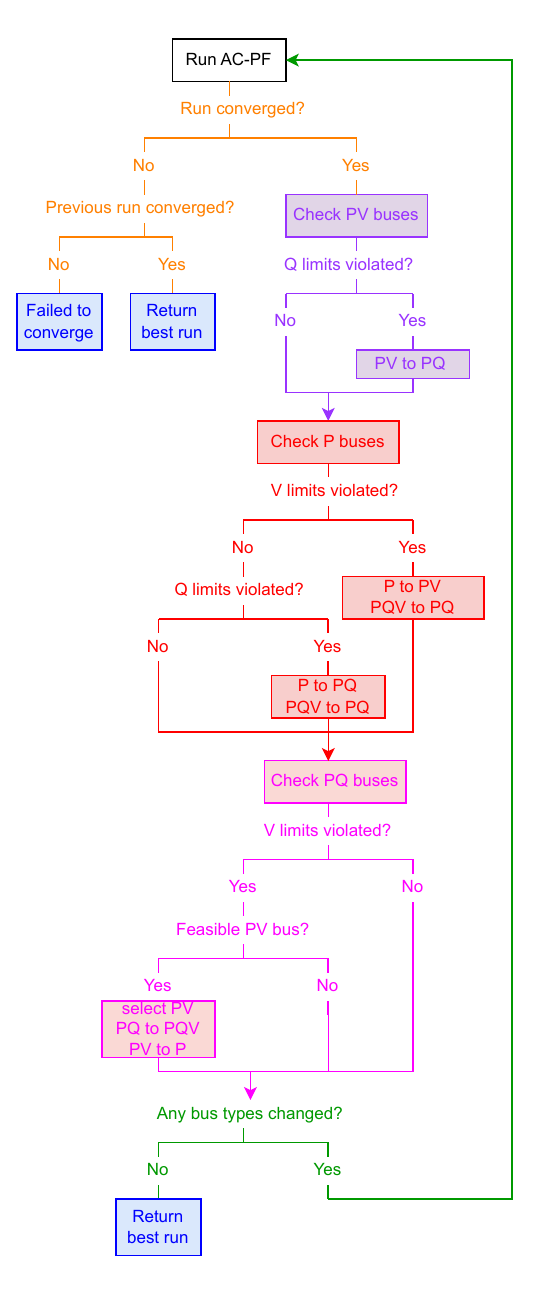} 
\caption{Proposed bus type switching algorithm.} \label{fig:flow}\vspace{-5mm} 
\end{figure}

\section{Methodology}\label{sec: methodology}
Our research is greatly inspired by two previous works. As discussed above, MATPOWER introduces an optional flag \textit{enforce\_q\_lims} that switches any PV bus with reactive power violations to a PQ bus \cite{matpower}. In \cite{bus_swap}, the authors discuss the convergence criteria required to model a system with both PQV buses and P buses. In this paper, we combine these two techniques. We propose a framework wherein the voltage magnitude at a PV bus is sacrificed in order to guarantee voltage stability at a load bus (PQV) elsewhere in the system. While other work has solved AC-PF with the inclusion of P and PQV buses, they typically represent physical system components where $P$, $Q$, and $V$ are controllable setpoints \cite{bs_remote_voltage, zbus_busswap}. Our framework does not assume the PQV buses have voltage control devices; it is simply a computational technique that will find a set of generator voltage magnitudes that do not cause voltage violations at load buses. To the best of our knowledge, this is the first paper to utilize bus type switching to resolve voltage constraint violations in AC-PF and achieve an AC-feasible setpoint.

\subsection{AC Power Flow with Bus Type Switching}\label{sec: idea}
Our proposed method includes two forms of bus type switching. In the first form (from \cite{matpower}), when reactive power at a PV bus is violated, the PV bus becomes a PQ bus and the NR-ACPF is re-run. In the second form, when voltage magnitude at a PQ bus is violated, the PQ bus becomes a PQV bus and a PV bus becomes a P bus. This form requires moving a degree of freedom to elsewhere in the system, and determining which PV bus to switch is a nontrivial problem. In \cite{bus_swap}, it is proved that to guarantee convergence of decoupled power flow with P and PQV buses, there must exist a unique path between each pair of P and PQV buses that does not contain a PV or V$\theta$ bus. These conditions are helpful, yet insufficient for fully coupled power flow; systems can still fail to converge when these conditions are met. Empirically, we found that the system is more likely to converge when the P and PQV bus are near one another. The first form of bus type switching has no limitations in the number of switches that can be made. In the second form, the number of violations that can be resolved is limited by the number and location of available PV buses.

\subsection{Flow Diagram}\label{sec: flow}
Figure \ref{fig:flow} depicts the flow diagram for our proposed technique. The first iteration will perform the baseline NR-ACPF described in Section \ref{sec: nracpf}. If any violations occur, the bus types at violated buses will be switched, and the NR-ACPF will be re-run. This framework prioritizes first minimizing voltage violations at generator buses, then minimizing reactive power violations at generator buses, and finally minimizing voltage violations at load buses. A violated PQ bus will only become PQV if there exists a feasible PV bus according to the criteria in \cite{bus_swap}. Finding unique paths from PQ to PV buses that do not pass through a PV or V$\theta$ bus is computationally expensive, so these paths are generated in advance for each test system. When a P-PQV pair is found, each bus on that path is removed from all other available paths. The PQ voltage violations are resolved in order from smallest to largest violation, and the PV bus with the shortest unique path to that PQ bus is selected. When an iteration fails to converge, the previous iteration with the smallest number of violations is returned. Therefore as long as the first iteration converges, our proposed method will return a solution that is equivalent or better than the baseline.

\begin{table*}[t!]
    \centering
    \begin{tabular}{|c|c|c|c|c|c|c|c|c|c|c|}
    \hline
       Test case  & ACPF & Feasible (\%) & \# Q viol. & \# V viol. & |V| viol. & \% \# V  & \% |V|  & Switch ratio & Avg.  Time (s) & Avg. Itrs \\
       \hline
       \multirow{4}{*}{case14} 
       & baseline & 4.5 & 1.85 & \multirow{2}{*}{2.73} & \multirow{2}{*}{0.38} &  \multirow{2}{*}{--} & \multirow{2}{*}{--}  & \multirow{2}{*}{--} &0.001 & 1  \\ 
       \cline{2-4} \cline{10-11}
       & qlim & 47.6 & \multirow{3}{*}{0} &  &  &   &  & & 0.001 & 1.77 \\ 
       \cline{2-3}\cline{5-11}
       & P-PQV & \multirow{2}{*}{54.2} &  & \multirow{2}{*}{1.1} & \multirow{2}{*}{0.007} & 30.9  & 39.1 &\multirow{2}{*}{3.82} & 0.002 & 3.71 \\ 
       \cline{2-2} \cline{7-8}\cline{10-11}
       & P-PQV' &  &  &  &  & 30.8  & 39.0  & & 0.002 & 3.85 \\
       \hline
       \multirow{4}{*}{case57} 
       & baseline & 0 & 3.69 & 1.5 & \multirow{2}{*}{0.012} &  \multirow{2}{*}{--}  & \multirow{2}{*}{--}  &\multirow{3}{*}{--} & 0.002 & 1  \\ 
       \cline{2-5} \cline{10-11}
       & qlim & 13.7 & \multirow{3}{*}{0} & 1.4 &  &   &  & & 0.003 & 2 \\
       \cline{2-3} \cline{5-8}\cline{10-11}
       & P-PQV & \multirow{2}{*}{72.5} &  & 0.77 & 0.011 & 30.0  & 30.0 & & 0.056 & 3.92 \\ 
       \cline{2-2}\cline{5-11}
       & P-PQV' &  &  & 0.57  & 0.009 & 45.0  & 40.3 & 1.0 & 0.056 & 4.17 \\
       \hline
        \multirow{4}{*}{case300} 
       & baseline & \multirow{4}{*}{0} & 9.52 & \multirow{2}{*}{5.77} & \multirow{2}{*}{0.054} & \multirow{2}{*}{--} & \multirow{2}{*}{--}& \multirow{3}{*}{--}& 0.014 & 1  \\ 
       \cline{2-2}\cline{4-4}\cline{10-11}
       & qlim &  & 0.77 &  &  & & & &0.021 & 2 \\
       \cline{2-2}\cline{4-8}\cline{10-11}
       & P-PQV &  & 0.8 & 4.7  & 0.048 & 24.8 & 18.6 & &1.16 & 6.07 \\ 
       \cline{2-2}\cline{4-11}
       & P-PQV' &  & 0.72 & 4.2 & 0.044 & 32.4 & 23.1 & 1.2 & 1.4 & 10.02 \\
       \hline
    \end{tabular}
    \caption{Statistics on AC power flow methods. \# Q and \# V show the average number of Q and V violations, |V| shows the average magnitude of V violation, \% \# V and \% |V| show the percentage improvement in the number and magnitude of V violations, and 'switch ratio' shows the ratio of P-PQV bus type switches to number of resolved V violations.}
    \label{tab:buswise_res}
\end{table*}

\section{Results}\label{sec: res}
Our method is tested on the IEEE 14-bus, 57-bus, and 300-bus test cases. We analyze results on 10,000 samples, where the real and reactive load is perturbed up to 85\% of the system's maximum available generation to simulate real-time market clearing with 15\% overall system generation headroom. Our method is compared to the baseline method seen in Section \ref{sec: nracpf} (baseline) and to the q-limit method seen in Section \ref{sec: qlim} (qlim). 
We show results for two implementations of our method. The first implementation (P-PQV) resolves all possible violations before re-running the NR-ACPF. This reduces the number of iterations, but resolves less violations. Alternatively, the second implementation (P-PQV') re-runs the NR-ACPF after each P-PQV bus type switch. 

\subsection{Violation Comparison by AC-PF Approach}\label{sec:acpfviol}
Table \ref{tab:buswise_res} shows results on the feasibility rate as well as number and magnitude of violations for each method. 
In the 14-bus test case, less than 5\% of samples are feasible without any bus type switching, and 54\% are feasible with the P-PQV approach.  This approach decreases the average number of voltage violations by 30\% and the average total magnitude of violations by 39\%. The switch ratio shows that on average, each P-PQV switch resolved voltage violations at 3.82 buses. A similar pattern can be seen in the 57-bus test case; with baseline AC-PF, none of the samples are feasible, while implementing P-PQV bus type switching restores feasibility in 72.5\% of samples. This approach decreases the average number of voltage violations by 30-45\% and total magnitude of voltage violations by 30-40\%.  In the 300-bus test case, all samples remain infeasible for every AC-PF approach. However, implementing bus-type switching reduces the number of reactive power violations by 92\% and the number of voltage violations by 24.8-32.4\%, with each P-PQV switch resolving voltage violations at more than one bus (a switch ratio greater than 1). 
Implementing the P-PQV bus type switching exponentially increases the convergence time and the number of iterations. While P-PQV' reduces a higher number and magnitude of violations compared to P-PQV, this comes with 4 more iterations on the 300-bus test system.

\begin{figure}[h!] 
\centering
\includegraphics[scale = 0.8, width = \columnwidth]{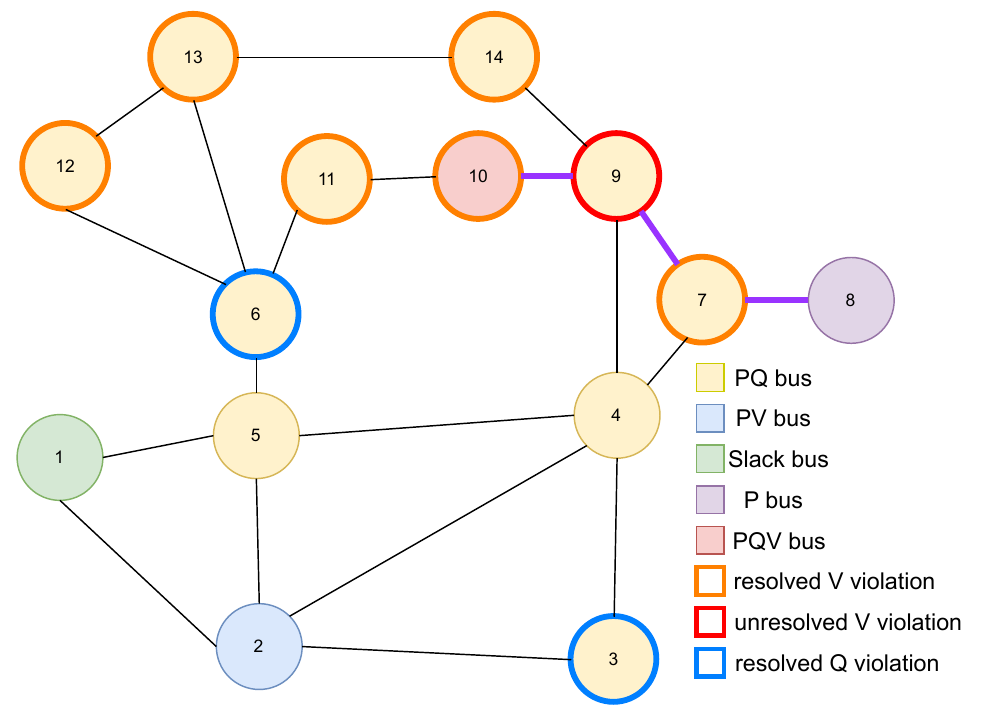} 
\caption{An example on the IEEE 14-bus test case. With normal NR-ACPF, 9 bound violations occurred after power flow. With the proposed algorithm, only one violation remains.} \label{fig:14bus}\vspace{-3mm} 
\end{figure}

\subsection{Example of the P-PQV Approach}\label{sec:14bus}
Figure \ref{fig:14bus} shows an example of the P-PQV bus switching approach on the 14-bus test case. With the baseline approach, this sample converged to a point where the reactive power limits were violated at 2 of the 4 PV buses (buses 3 and 6) and voltage limits were violated at 7 of the 9 PQ buses (bus 7 and buses 9 through 14). With bus type switching, the 2 violated PV buses became PQ buses. Bus 10 and bus 8 became a P-PQV pair, connected by the path highlighted in purple.  Although the voltage magnitude at bus 9 was still violated, the rest of the voltage and reactive power violations were resolved. 

The voltage setpoints for both the baseline approach and the P-PQV approach for the 14-bus sample are displayed in Table \ref{tab:14bus}. Sacrificing the voltage setpoints at buses 3 and 6 only changed the voltage magnitudes by 0.003 V and 0.009 V, respectively. Reducing the voltage setpoint at bus 8 by 0.031 V caused the voltage setpoints at buses 7 and 10 through 14 to fall within their boundaries. In the baseline approach, the voltage magnitude at bus 9 violated its upper bound by 0.018 V; after sacrificing the voltage setpoint at bus 8, bus 9 only violated the upper bound by 0.004 V.

\begin{table}[h!]
    \centering
    \begin{tabular}{|c|c|c|c|c|c|}
    \hline
     & \textbf{Baseline} & \textbf{P-PQV} & & \textbf{Baseline} & \textbf{P-PQV} \\ 
     \hline
     $V_3$ & 1.01 & 1.001 & $Q_3$ & \textcolor{red}{-.107} & 0 \\ 
     \hline
     $V_6$ & 1.07 & 1.061 & $Q_6$ &\textcolor{red}{-.106} & -.06 \\
     \hline
    $V_7$ & \textcolor{red}{1.076} & 1.059 & $V_8$ & 1.09 & 1.059 \\
    \hline
    $V_9$ & \textcolor{red}{1.078} & \textcolor{red}{1.064} & $V_{10}$ & \textcolor{red}{1.073}  & 1.06 \\
    \hline
    $V_{11}$ & \textcolor{red}{1.0704}  & 1.059 & $V_{12}$ & \textcolor{red}{1.064}  & 1.055 \\ 
    \hline 
    $V_{13}$ & \textcolor{red}{1.063}  & 1.053 & $V_{14}$ & \textcolor{red}{1.063}  & 1.051\\
    \hline
    \end{tabular}
    \caption{Numerical values for the aforementioned 14-bus test case. Violations highlighted in red.}
    \label{tab:14bus}
\end{table}

\subsection{Voltage Violation Distribution}\label{sec:300bus}
\begin{figure}[h!]
    \centering

    % --- Top to bottom stack ---
    \begin{subfigure}[t]{0.4\textwidth}
        \centering
        \includegraphics[width=\textwidth]{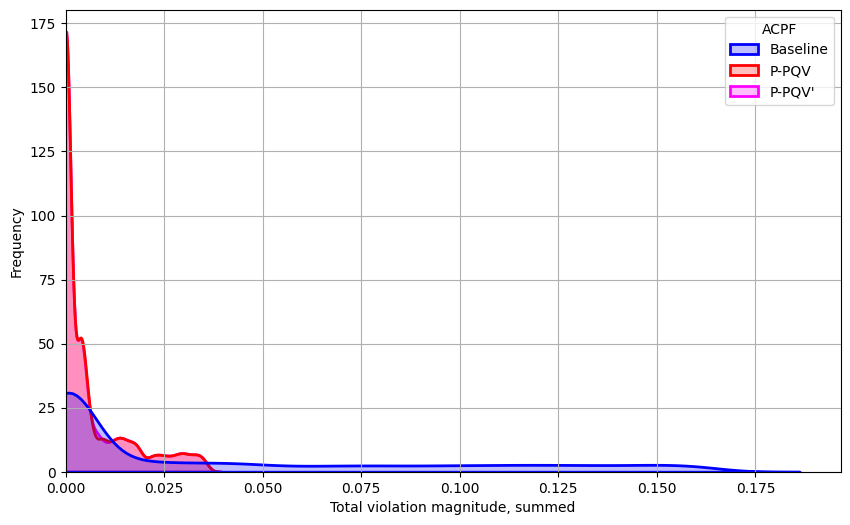}
        \caption{14-bus distribution.}
    \end{subfigure}
    \vspace{2mm}
    \begin{subfigure}[t]{0.4\textwidth}
        \centering
        \includegraphics[width=\textwidth]{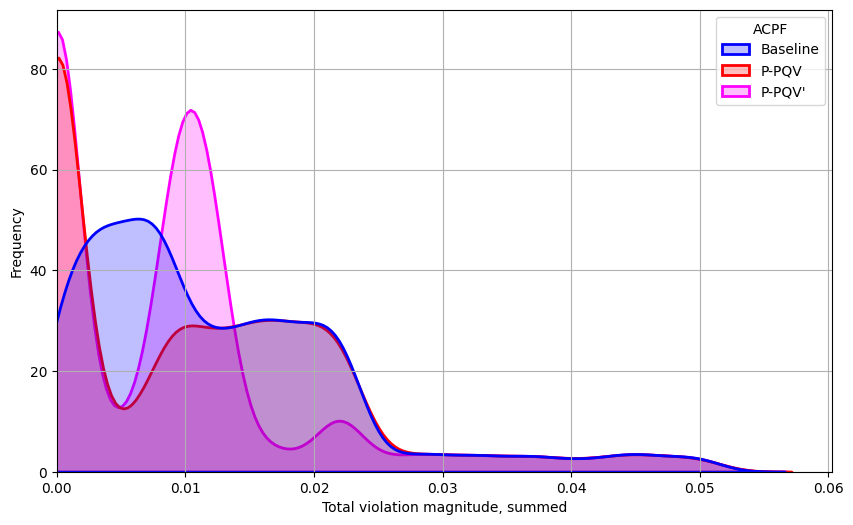}
        \caption{57-bus distribution.}
    \end{subfigure}
    \vspace{2mm}
    \begin{subfigure}[t]{0.4\textwidth}
        \centering
        \includegraphics[width=\textwidth]{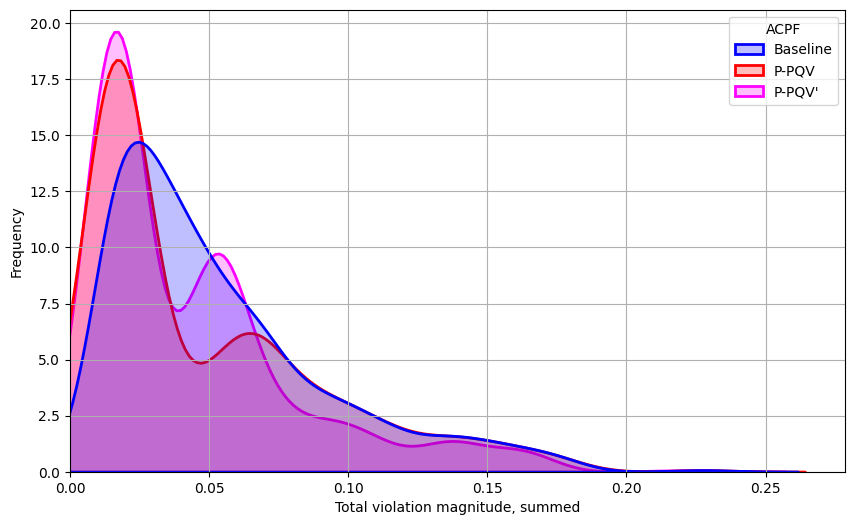}
        \caption{300-bus distribution.}
    \end{subfigure}

    \caption{Distribution of the total magnitude of voltage violations in the IEEE 14-bus, 57-bus and 300-bus test cases.}
    \label{fig:distributions}
\end{figure}

Figure \ref{fig:distributions} shows the distribution of total voltage magnitude violation in each test case under various AC-PF implementations. In the 14-bus test case, the magnitude violation with the P-PQV and P-PQV' approach is more tightly concentrated around 0, without the large tail seen in the baseline distribution. The 57-bus and 300-bus test case have different distributions between the P-PQV and P-PQV' approaches. Both distributions show again a higher concentration around 0 compared to the baseline distribution. However, the P-PQV' distribution has a second peak to the left of the baseline distribution, while the P-PQV approach either has the minimal violation, or the same violation as the baseline. Both of these approaches have the same long tail as the baseline, as well.

% \begin{figure*}[h!]
%     \centering
%     % --- Top row ---
%     \begin{subfigure}[t]{0.3\textwidth}
%         \centering
%         \includegraphics[width=\textwidth]{figs/57_vmnumberdist.png}
%         \label{fig:57_num}
%     \end{subfigure}
%     \begin{subfigure}[t]{0.3\textwidth}
%         \centering
%         \includegraphics[width=\textwidth]{figs/57_vmdist.png}
%         \label{fig:57_viol}
%     \end{subfigure}
    
%     \vspace{3mm} % space between rows
    
%     % --- Bottom row ---
%     \begin{subfigure}[t]{0.3\textwidth}
%         \centering
%         \includegraphics[width=\textwidth]{figs/300_vmnumberdist.png}
%         \label{fig:300_num}
%     \end{subfigure}
%     \begin{subfigure}[t]{0.3\textwidth}
%         \centering
%         \includegraphics[width=\textwidth]{figs/300_vmdist.png}
%         \label{fig:300_viol}
%     \end{subfigure}
%     \caption{Distribution of number and magnitude of VM violations in the IEEE 57-bus and 300-bus test cases.}
%     \label{fig:distributions}
% \end{figure*}

\section{Discussion and Conclusion}\label{sec: disc}
We presented results for our proposed method of AC-PF feasibility adjustment on the IEEE 14-bus, 57-bus, and 300-bus test cases. Across all test cases, our method reduced both the average number of reactive power violations and the average number of voltage  violations, often resulting in a completely feasible power flow solution. Because voltage is localized in power systems, by resolving a voltage violation at one bus, our method can reduce the number and total magnitude of voltage violations at neighboring buses.

The primary disadvantage of our method is that it requires additional iterations of NR-ACPF, and finding suitable P-PQV pairs can be computationally expensive. That said, since the first iteration of our method performs the baseline NR-ACPF, imposing a maximum iteration cutoff will still return a solution with equal or less total violations than the baseline approach. 

Several aspects of our method should be further explored. The order in which P-PQV pairs are selected matters, since the path between each pair limits the remaining feasible paths. For instance, in the 14-bus sample shown in Figure \ref{fig:14bus}, bus 10 gained a voltage setpoint from bus 8. Since bus 9 is on the path between these buses, its bus type cannot be switched. If bus 10 had instead gained a setpoint from bus 2 (a feasible path, since bus 6 became a PQ bus), bus 9 may have paired with bus 8, instead. Our method also does not consider the interactions between connected buses. As seen by the switch ratio for the 14-bus and 300-bus test case,resolving a voltage violation at one bus can also resolve violations at neighboring buses. However, adjusting a voltage setpoint can also have the opposite effect and push the voltage or reactive power at neighboring buses past their boundaries. In sum, the method presented here is not guaranteed to produce the setpoints with the minimum number and magnitude of violations. 

Importantly and unlike many other works which require generator dispatch changes in order to attain AC feasibility, this technique does not deviate from the setpoints cleared in the market, nor does it require shedding load or adjusting physical system components. Moreover, as this technique is not data-driven or a black box model, the results are interpretable and reliable. While the work introduced in this paper is promising, there are many opportunities to expand upon it. Our work does not consider violations in thermal limits or the interactions between buses, and its practical utility is reduced due to the increased number of AC-PF iterations. Future work should look to further reduce the number of system violations without increasing computation time.

\bibliographystyle{IEEEtran}
\bibliography{references}
\end{document}